\documentclass[%
 reprint,
 superscriptaddress,
 amsmath,amssymb,
]{revtex4-1}

\usepackage{graphicx}
\usepackage{dcolumn}
\usepackage{bm}
\usepackage{hyperref}
\usepackage{url}
\usepackage{comment}
\usepackage{xspace}
\usepackage{mathtools}
\usepackage{tabularx}

\newcommand{\fref}[1]{Fig.~\ref{#1}}
\newcommand{\Fref}[1]{Figure~\ref{#1}}
\newcommand{\tref}[1]{Table~\ref{#1}}

\newcommand{\appropto}{\mathrel{\vcenter{
  \offinterlineskip\halign{\hfil$##$\cr
    \propto\cr\noalign{\kern2pt}\sim\cr\noalign{\kern-2pt}}}}}

\newcommand{\tablecell}[1]{\parbox{12cm}{\vspace{0.5em}\raggedright #1\vspace{0.5em}}}

\newcommand{\mycomment}[1]{}

\begin{document}
\newcommand{\ManuscriptTitle}{
    Application of the Portable Diagnostic Package to the Wisconsin High-temperature-superconducting Axisymmetric Mirror (WHAM)
}

\title{\ManuscriptTitle}

\author{Keisuke Fujii}
\email{fujiik@ornl.gov}
\affiliation{%
    Fusion Energy Division, Oak Ridge National Laboratory, Oak Ridge, TN 37831-6305, United States of America
}
\author{Douglass Endrizzi}
\affiliation{%
    Realta Fusion Inc., Madison, WI 53717, United States of America
}
\author{Jay K. Anderson}
\affiliation{%
    UW-Madison Department of Physics, University of Wisconsin Madison, Madison, WI 53706, United States of America
}
\author{Cary B. Forest}
\affiliation{%
    UW-Madison Department of Physics, University of Wisconsin Madison, Madison, WI 53706, United States of America
} 
\author{Jonathan Pizzo}
\affiliation{%
    UW-Madison Department of Physics, University of Wisconsin Madison, Madison, WI 53706, United States of America
}
\author{Tony Qian}
\affiliation{%
    Princeton University, Princeton, NJ 08544, United States of America
}
\author{Mason Yu}
\affiliation{%
    UW-Madison Department of Physics, University of Wisconsin Madison, Madison, WI 53706, United States of America
}
\author{Theodore M. Biewer}
\affiliation{%
    Fusion Energy Division, Oak Ridge National Laboratory, Oak Ridge, TN 37831-6305, United States of America
}

\date{\today}

\begin{abstract}
    We present an application of the Portable Diagnostic Package (PDP) on the Wisconsin High-temperature-superconducting Axisymmetric Mirror (WHAM), which integrates an optical emission spectroscopy (OES) system and an active Thomson scattering (TS) system.
    The OES system facilitates a comprehensive impurity line survey and enables flow measurements through the Doppler effect observed on impurity lines. 
    Plasma rotation profiles were successfully derived from doubly charged carbon lines.
    The TS system enabled the first measurements of the electron temperature in commissioning plasmas on WHAM.  
    Notably, the PDP was installed, commissioned, and used to obtain OES and TS data within $\sim$\,6 months from project start, enabled by its designed portability and standardized interfaces. These results demonstrate the PDP's potential to accelerate diagnostic readiness and advance experimental plasma studies.
\end{abstract}

\maketitle
\onecolumngrid

\section{Introduction}

Magnetic confinement devices and high-temperature plasmas serve as key platforms for testing plasma physics theories and advancing fusion research. Accurate measurements of plasma parameters—such as electron temperature, density, plasma rotation, and impurity content—are essential for understanding transport processes and for optimizing operational regimes. Over the years, diagnostic methods including passive optical emission spectroscopy (OES) and active Thomson scattering (TS) have proven invaluable for these tasks. However, while the underlying diagnostic principles are well established, the ability to quickly integrate these systems into different experimental setups has remained a challenge.

To facilitate rapid validation of emerging confinement device concepts, the Advanced Research Projects Agency - Energy (ARPA-E) ``Topics Informing New program Areas'' (TINA) program established several diagnostic capability teams and funded portable diagnostic systems~\cite{osti_1992496, osti_1848375, wurden2021portable, banasek2023probing}.  
Oak Ridge National Laboratory (ORNL) was established as one of several diagnostic capability teams. 
Our portable diagnostic package (PDP) combines an OES system and an active TS system into a single, compact unit that can be rapidly deployed and integrated into existing plasma experiments~\cite{kafle2021design}. 
Unlike conventional OES and TS setups that are typically fixed and require extensive reconfiguration, our PDP emphasizes portability and ease of integration.  This concept was detailed in our earlier work ~\cite{he2022implementation, kafle2022portable}.
The OES system serves to quickly identify impurity species and assess plasma flow via Doppler-shift analysis.
Concurrently, TS measurements in our package use a pulsed laser to illuminate the plasma and extract localized electron temperature and density profiles.
In this work, we report the application of the PDP to commissioning plasmas on the Wisconsin High-Temperature-Superconducting Axisymmetric Mirror (WHAM)~\cite{endrizzi2023physics} under a 2023 Innovation Network for Fusion Energy (INFUSE) project. 

The remainder of this paper is organized as follows. Sec.~\ref{sec:setup} describes the experimental setup and the architecture of the integrated PDP. Sec.~\ref{sec:results} outlines the experimental results from WHAM. Finally, Sec.~\ref{sec:summary} summarizes our findings and proposes directions for future improvement.

\section{Experimental Setup\label{sec:setup}}

\subsection{WHAM}

The WHAM experiment is designed to demonstrate the performance of and investigate the physics of the high-field axisymmetric magnetic mirror. 
This device is engineered to bridge the gap between gas-dynamic regimes and classical mirror confinement by leveraging high-temperature superconducting (HTS) magnets to achieve large mirror ratios and enhanced electron thermal confinement~\cite{endrizzi2023physics, Soldatkina_2017}. 

\begin{figure}[tb]
    \includegraphics[width=17cm]{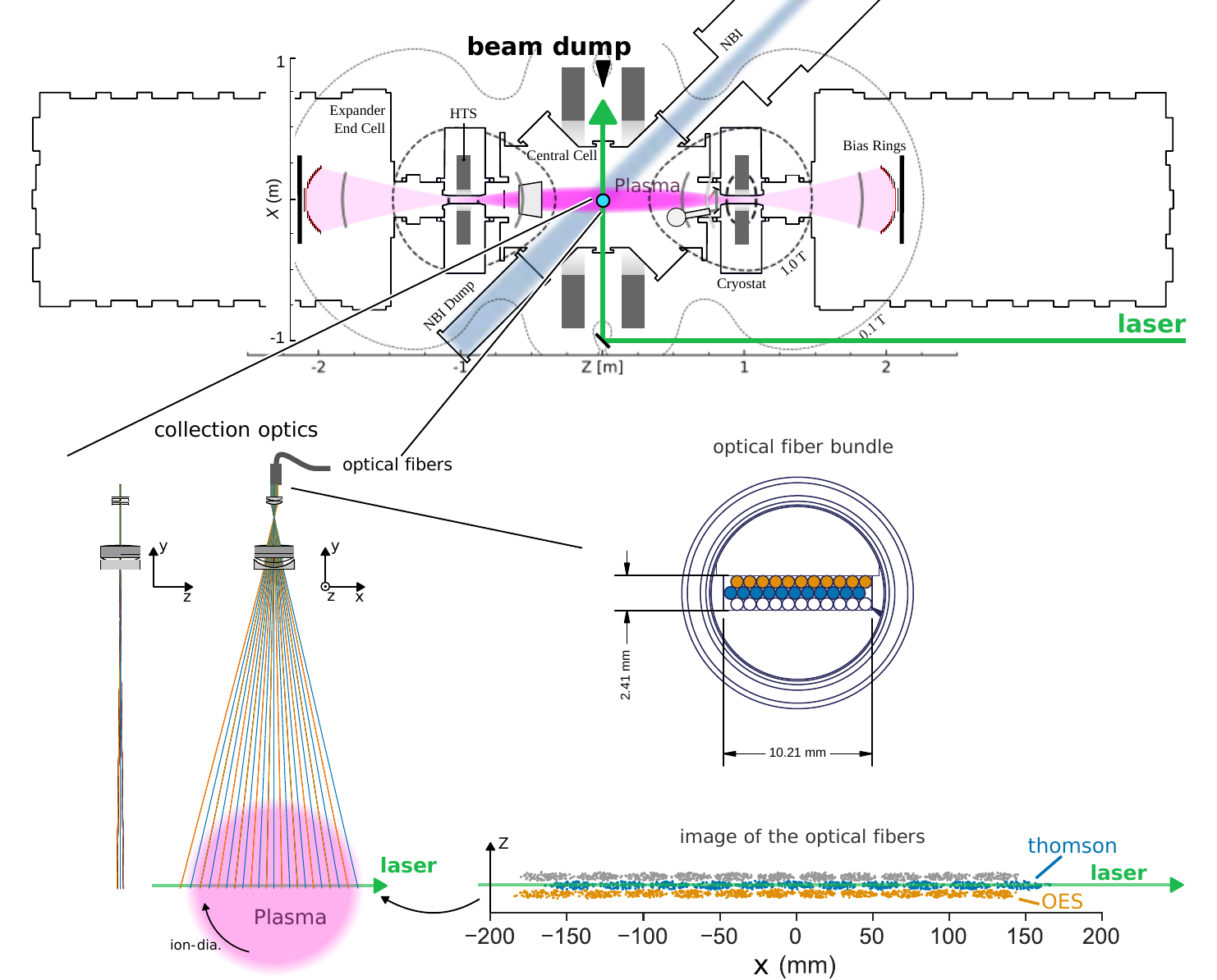}
    \caption{%
        A schematic illustration of the measurement system.
        (top) Overview of WHAM: The laser is injected into the central cell on the midplane. 
        The collection optics for the TS and OES systems are attached on the top port.
        (lower left) Collection optics: The optics consist of a 75\,mm achromat and multiple cylindrical lenses, to expand the image along the laser.
        (middle right) Optical fiber bundle: The light is focused on 3$\times$11 fibers, which are in the bundle. 
        The central row is used for the TS and the bottom row is used for the OES.
        (bottom right) Image of the optical fibers on the midplane, simulated by an optical ray tracing program. 
        The image is expanded along the laser direction by a factor of $\sim$\,5. 
        The image size of each fiber is $\sim$\,13\,mm in the laser direction and $\sim$\,2.7\,mm across the laser path. 
    }
    \label{fig:configuration}
\end{figure}

A schematic illustration of WHAM is shown in \fref{fig:configuration}~(top).
The device employs 17\,T HTS mirror magnets custom-built by Commonwealth Fusion Systems. 
A gyrotron-based electron cyclotron heating (ECH) system operating at 110\,GHz is used to ionize the target plasma, while a 25\,keV, 40\,A deuterium neutral beam (NB) generates a sloshing ion distribution essential for sustaining the desired confinement regime. 
In addition, a biased limiter and end-ring system is implemented to impose an edge electric field that stabilizes magnetohydrodynamic interchange modes through sheared $E\times B$ drifts and maintains high-performance plasma conditions~\cite{Ryutov_2011,Bagryansky_2007}. 

\subsection{PDP Overview and Optical Configuration}
\label{sec:pdp-overview}

Developed at ORNL, the PDP is a compact, rapidly deployable system that integrates OES and TS for comprehensive plasma diagnostics. 
A pulsed, frequency-doubled Nd:YAG laser (Lumibird Q-SMART 1500) drives the TS module, producing 6\,ns pulses at 532\,nm with energies up to 800\,mJ at a 10\,Hz repetition rate. The laser beam diameter is approximately 10\,mm and is injected on the midplane of WHAM along the $X$ axis (see \fref{fig:configuration}). A 2\,m focal-length lens focuses the beam to a spot size of less than 1\,mm at the center of the plasma. After passing through the plasma, the laser is absorbed by an \textit{ex vessel} atmospheric beam dump (Acktar LBD-T-30), which is rated with a $10^{-6}$ reflectivity factor. 

The collection optics are shared by both the TS and OES systems.
We employ an existing optical fiber bundle (3$\times$11 fibers, each with 800\,$\mu$m core diameter, 880\,$\mu$m cladding diameter, and NA=0.12; see \fref{fig:configuration}\,(center\,right)).
We have designed new optics primarily from commercial off-the-shelf components, 
to accommodate the large disparity between the $\sim$\,300\,mm diameter plasma and the $\sim$\,1\,mm laser beam, while maximizing the scattered-light collection efficiency. 

A schematic of the collection system is shown in the bottom-left of \fref{fig:configuration}. The design uses multiple lenses: a concave cylindrical lens (Thorlabs LK1431L1, 75\,mm focal length, 53.0$\times$50.8\,mm, N-BK7), an achromatic lens (Thorlabs AC508-075-A, 75\,mm focal length, 50.8\,mm diameter) and two convex cylindrical lenses (Thorlabs LJ1125L1-A, 40\,mm focal length, 20$\times$22\,mm, N-BK7). These are mounted in a Thorlabs 60\,mm cage system with 3D-printed holders. Optical ray tracing with BeamFour (Stellar Software) confirms that the cylindrical lenses provide different magnifications along $X$ and $Z$, resulting in a $\sim$\,$13\times 2.7$\,mm image at the midplane for each 0.8\,mm fiber.
See the bottom-right of \fref{fig:configuration} for the simulated image by the ray-tracing.
As described later, the smaller image size across the laser beam increases the light collection efficiency by a factor of $\sim$\,5 compared with an optics covering the same plasma size but with the identical magnifications in $X$ and $Z$ directions.

Light collected by these optics is transferred via optical fibers to the PDP spectrometers, located about 5\,m away. Since our existing fiber bundle is only 4\,m long, we extended it using 600\,$\mu$m diameter, NA=0.22 fibers from our inventory, which unfortunately reduced the throughput by more than 50\%. 
The fibers are then reconnected to 800\,$\mu$m diameter, NA=0.12 fibers leading into two Teledyne Princeton Instruments Isoplane320 spectrometers, one for TS and one for OES.
The entrance slit for the TS spectrometer is set to be 200 $\mu$m, while 30 $\mu$m is employed for the OES spectrometer.
The wavelength and the instrumental profiles were calibrated by using a neon lamp.

Each spectrometer is equipped with three gratings. 
The TS spectrometer uses 300/mm, 600/mm and 1200/mm gratings (all blazed at 500\,nm). 
The OES spectrometer is equipped with 150/mm, 1800/mm and 2400/mm gratings to provide both wide wavelength coverage and high-resolution capabilities. 
Detection is performed by a PI-MAX 4-1024f image-intensified CCD camera (1024$\times$1024 pixels, 13\,$\mu$m pixel size). 
This camera can be gated as short as 500\,ps, effectively suppressing background plasma emission during TS measurements.


\subsection{Optical Emission Spectroscopy}
\label{sec:oes}

The OES diagnostic is used both to survey impurities and to measure the plasma rotation in WHAM. The wavelength range and grating selection are adjusted shot by shot, according to experimental requirements. The exposure time is controlled by the image intensifier. Due to the camera's readout duration (lasting $\sim$\,15\,ms), only one data-frame can be acquired per discharge. This timing is adjusted in software, and a typical exposure is on the order of $\sim$\,0.1\,ms, depending on plasma conditions.

\begin{figure}[tb]
    \includegraphics[width=17cm]{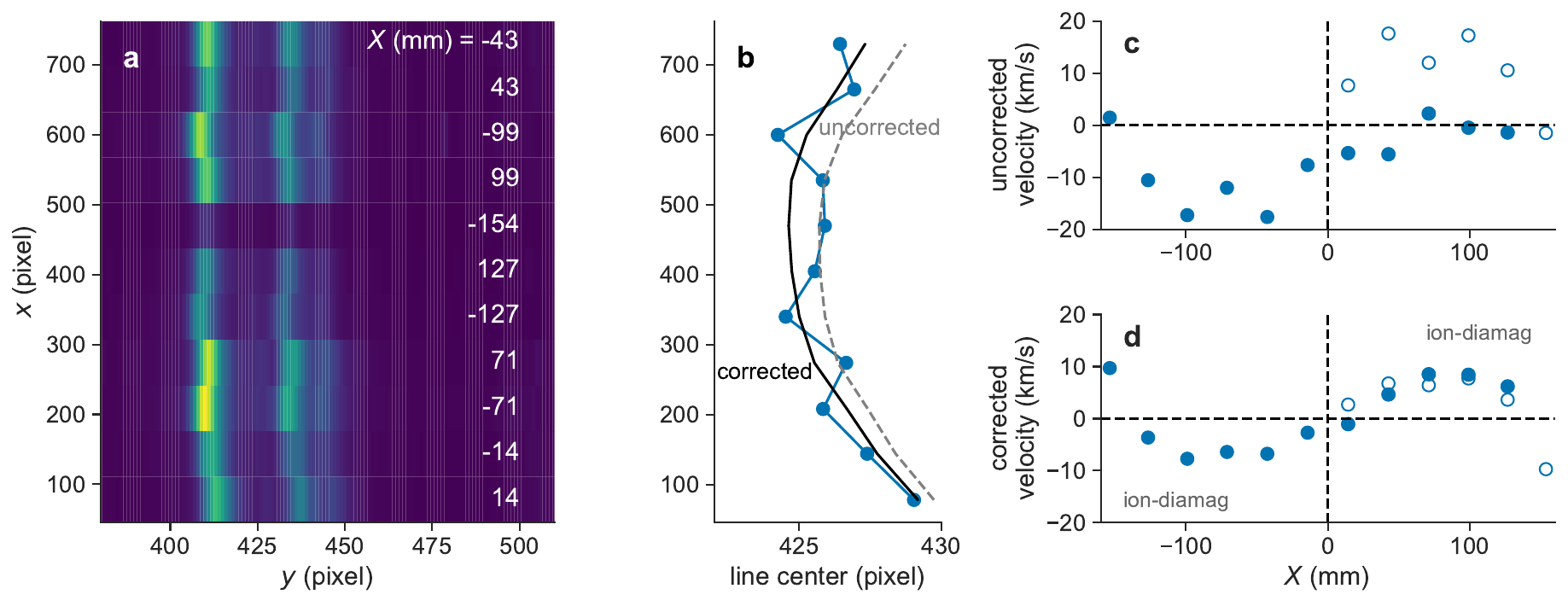}
    \caption{%
    (a) CCD image of the doubly charged carbon emission lines, with the horizontal $y$ axis corresponding to the wavelength-dispersion direction and the vertical $x$ axis parallel to the entrance slit. The labeled $X$-positions denote the chords associated with each optical fiber.
    (b) Extracted line-center positions from each chord (blue markers). The dashed line indicates the line center position based on the nominal calibration, while the solid black line shows the best-fit shift that yields a more symmetric velocity profile.
    (c) Plasma velocity profile obtained from the line shift as a function of the distance from the plasma axis, $X$. 
    Open markers are the velocity observed in $X < 0$ cast to the positive $X$-side assuming axi-symmetry of the plasma.
    The velocity profile using the nominal calibration shows strong asymmetry, i.e., the velocity values obtained in the corresponding positions (e.g., +43 mm and -43 mm) are largely different.
    (d) Plasma velocity profile after applying the correction derived from (b), now displaying a more realistic, near-symmetric shape about the plasma center.
    }
    \label{fig:OES_image}
\end{figure}

Measuring plasma rotation requires sub-pixel resolution, even with a 2400\,/mm grating, so small drifts in the spectrometer hardware (e.g. mechanical shifts or thermal expansion) can undermine wavelength calibration. To mitigate these effects, we implemented an alternating fiber configuration, whereby fibers viewing opposite sides of the plasma are placed adjacent to one another. \Fref{fig:OES_image}\,(a) shows a CCD image of doubly charged carbon emission lines, with the $X$-axis location in the plasma for each view labeled. The paired arrangement of fibers ensures that any global calibration shift can be distinguished from genuine rotation effects.

\Fref{fig:OES_image}\,(b) illustrates the extracted line centers for each fiber (see Sec.\ref{sec:results} for the line-fitting procedure). The measured positions lie systematically to one side of the nominal wavelength center (gray dotted curve in \fref{fig:OES_image}\,(b)), which produces a strongly asymmetric velocity profile across the plasma (\fref{fig:OES_image}\,(c)). Because the plasma is expected to rotate rather than translate, we interpret this asymmetry as an artifact arising from spectrometer misalignment or calibration drift.

To correct for this offset, we assume that the plasma is rotating while the spectrometer introduces a finite horizontal shift and tilt. We then adjust these parameters so that the resulting velocity profile becomes maximally symmetric about the plasma center. The black solid line in \fref{fig:OES_image}\,(b) denotes the adjusted wavelength center, and the resulting velocity profile is shown in \fref{fig:OES_image}\,(d). This robust calibration procedure relies on our alternating fiber arrangement; if the fibers were arranged sequentially, it would be impossible to decouple detector tilt from genuine plasma rotation.

\subsection{Thomson Scattering}
\label{sec:ts}

One of the major challenges in TS diagnostics is mitigating stray light from the laser. Because the Thomson scattered-signal intensity can be smaller than that of the laser by a factor of $\sim$\,$10^9$, even moderate reflections or beam halos can overwhelm the TS spectrum. Although the main beam is dumped in the dedicated beam dump ($10^{-6}$ reflectivity factor), strong stray light can still arise from multiple reflections in-vessel.

A significant source of stray light in our system is the extended halo of the primary laser beam. 
While the main beam diameter is about 10\,mm, its far wings strike the in-vessel walls, especially around the laser exit port opening, and introduce additional stray light. 
Light-trapping materials for shallow angle incidence (Acktar Ltd., Hexa-Black rolled-panels) were added to the inner surfaces of the laser inlet and outlet beam tubes.  
To mitigate reflected stray light, we installed an optical backstop (Acktar Ltd., Spectral Black coated foil without adhesive) on the vacuum-chamber wall opposite the collection optics, reducing the stray light intensity by approximately an order of magnitude. 
Note that due to the limitation of the vacuum vessel geometry, only the central five sight lines terminate on the optical backstop. The other sight lines suffer from strong stray light and are currently not-useable for TS measurements.
Additionally, the asymmetric magnification in the $X$ and $Y$ directions of the collection optics improves the scattered-light collection efficiency by a factor of $\sim$\,5, since the bigger magnification along $Z$ direction enables a bigger aperture from the scattered volume.
On the other hand, a similar stray-light background level is maintained, because the bigger aperture for the stray light is cancelled out by the smaller viewing volume along $Z$ direction.

Even though the TS signal is shifted in wavelength relative to the laser line, strong stray light can still affect other parts of the spectrum in our detection system. Image intensifiers often exhibit a finite cross-talk, where intense input signals can bleed over into nearby pixels~\cite{Ientilucci1996-rv,Xu2025-sq}, thus degrading TS measurements. Also, the strong stray-light risks damage to the intensifier. 

\begin{figure}[tbp]
    \centering
    \includegraphics[width=0.8\linewidth]{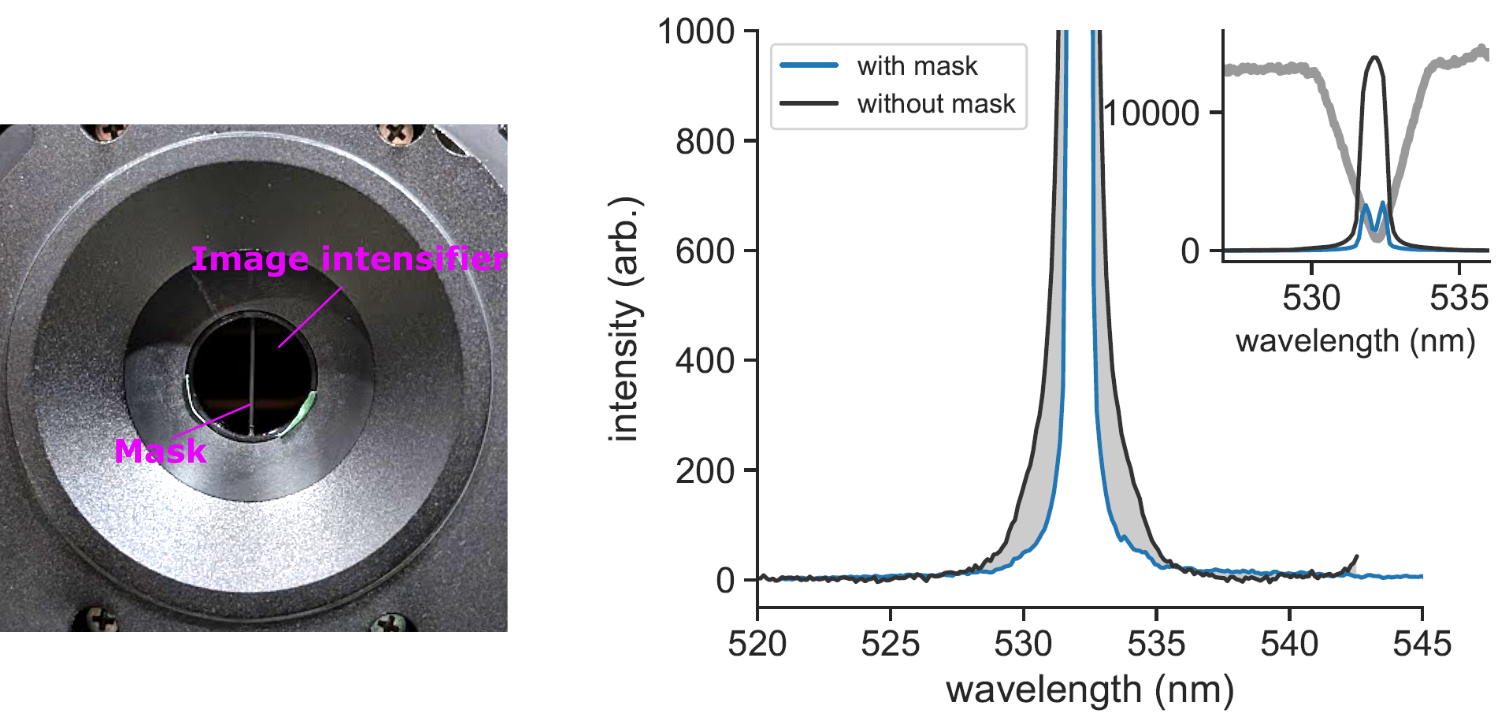}
    \caption{%
    (a) Photograph of the image intensifier with the 3D-printed optical mask installed. 
    The central opening is $\sim$\,0.6\,mm in diameter and mounted off-focus to reduce 
    the intense stray-light halo. 
    (b) Measured stray-light spectra without (black) and with (blue) the mask, showing 
    an overall five-fold reduction and substantially decreased wings caused by intensifier 
    cross-talk. The inset (gray) depicts the mask's triangular transmission profile 
    across the detector plane.    
}
    \label{fig:TS_mask}
\end{figure}

To address this, we installed a 3D-printed optical mask (made of black PolyLactic Acid (PLA, 0.6\,mm width) in front of the input window of the intensifier, as shown in \fref{fig:TS_mask}\,(a). 
Since the spectrometer focus is on the photocathode of the intensifier, which is coated on the backside of the input window, the mask is not exactly on the focus. 
Although the camera manufacturer does not disclose the input-window thickness, this might be $\sim$1\,mm. 
This configuration results in $\sim$0.2\,mm defocus of the mask, which is a similar size to the entrance slit width.
The inset of \fref{fig:TS_mask}\,(b) (gray curve) illustrates the triangular transmission profile introduced by this off-focus placement, with a minimum transmission of $\sim$\,$10^{-2}$ at the center. Figure~\ref{fig:TS_mask}\,(b) compares stray-light spectra acquired without (black) and with (blue) the mask, showing a factor-of-five reduction in total stray light and a significantly diminished wing component, which arises mainly from intensifier cross-talk.
Note that some fraction of the light hitting the optical mask will not be absorbed but scattered, which may result in another source of stray light. 
However, even with this effect, the total stray light is significantly reduced by the optical mask (as can be seen in \fref{fig:TS_mask}\,(b)), and will be in principle subtracted by the stray-light subtraction procedure described below.

Since currently the pulse length of WHAM experiments is limited by the power supply to $\sim$\,15\,ms, and our laser repetition frequency is 10 Hz (100\,ms period), we can only measure a single laser pulse during each plasma discharge.
Two camera exposures are made for each experiment; one exposure during the plasma discharge, the second exposure (100\,ms later) during the laser injection after the discharge has ended.
The first exposure captures the Thomson scattering light, the stray light, and the emission from the plasma, while the second exposure captures only the stray light component.  These are subtracted to remove the stray light contribution, without correcting for the variance in each laser pulse. 
The emission from the plasma also affects the signal to noise ratio.
To minimize the background emission, we use a 20\,ns exposure time, which is close to the pulse width of the laser (6\,ns), while still accommodating modest timing-jitter.
This timing sequence is summarized in Appendix~\ref{app:timing}.

As shown later, the background emission from WHAM still affects the TS measurement. To discriminate the plasma emission from the TS spectrum while maintaining the photon throughput, we use a 200 $\mu$m slit width and 1200/mm grating for the TS spectrometer.
The wavelength resolution of the TS system is $\sim$\,0.8\,nm FWHM.

Finally, the TS system's optical sensitivity is calibrated via Rayleigh scattering measurements, as discussed in Appendix~\ref{app:RayleighCalibration}. 

\subsection{Timeline of the PDP installation}\label{app:sec:timeline}

\begin{table}[tbp]
    \centering
    \renewcommand{\arraystretch}{1.2}
    \caption{Summary of Visits and Key Activities}
    \begin{tabular}{c c c}
    \hline
    \textbf{Visit} & \textbf{\;\;Project month} & \textbf{Activity Summary} \\ 
    \hline
    -- & PM1 & \tablecell{
        Discussion and planning.
    } \\ 
    \hline
    1 & PM2 & \tablecell{
        Transport of PDP spectroscopy system from ORNL to WHAM and installation on device, integrated with WHAM timing. \textbf{First OES measurements from WHAM plasmas were achieved}.
    } \\ \hline
    2 & PM3 & \tablecell{Transport of PDP laser system from ORNL to WHAM and installation in machine area. \textbf{OES analysis automated for routine operation}.} \\ \hline
    3 & PM4 & \tablecell{Laser beam line enclosures established and laser-safety approval petitioned. TS laser integrated with WHAM timing.} \\ \hline
    4 & PM5 & \tablecell{Open-beam laser operation approved, collection optics aligned, and stray light assessed.} \\ \hline
    5 & PM6 & \tablecell{Stray light improvements implemented and TS attempt on plasma (unsuccessful).} \\ \hline
    -- & PM7 & \tablecell{No visit due to scheduling conflicts.} \\ \hline
    6 & PM8 & \tablecell{Stray light and timing improvements implemented. \textbf{TS attempt on plasma successful}; two experimental conditions measured as reported here.} \\ \hline
    \end{tabular}
    \label{tab:visit_summary}
\end{table}

The PDP was designed for rapid deployment.  The ARPA-E TINA program that established such diagnostic capability teams used metrics of performance that included "time to initial deployment" and "time between deployments."  This paper describes the third deployment of the PDP within 5 years, including initial deployment on the Electro-thermal arc (ET-arc) plasma source at ORNL~\cite{he2022implementation} and the Princeton Field Reversed Configuration-2 (PFRC2) at Princeton Plasma Physics Laboratory (PPPL)~\cite{kafle2022portable}.
All components---two spectrometers, cameras and controllers, a control computer with monitors, optical fiber bundles, and fiber patch panels---are integrated into a single cart (with the exception of the laser, which is mounted on a separate cart) that fits into a commercial van.  The inventory of PDP equipment weighs $\sim$900 pounds.
This portable configuration enabled the successful deployment on WHAM of both the OES and TS systems within eight months, by two researchers working intermittently.  As a historical point of reference, when in 1969 a UKAEA team deployed a Thomson scattering system to the T3 tokamak in Russia, the shipment of 26 crates weighing $\sim$5\,tons was transported by cargo plane.~\cite{forrest}  After 3 months of planning, that TS system was deployed in $\sim$6 months by a team of 4 full-time researchers (including some credited with the invention of TS~\cite{DeSILVA1964-gp}), producing measurements that profoundly influenced the course of fusion energy research in the world community.~\cite{Peacock1969-pf}.  The T3 measurements often appear as the first plotted point for tokamak fusion triple-product in Lawson criterion plots, such as Fig 3 in ~\cite{Peacock1969-pf}.

A brief timeline of the installation of the PDP on WHAM is given in \tref{tab:visit_summary}, with major tasks described:
The first Project Month (PM1) consisted of discussions and planning.  Subsequently, ORNL personnel accomplished installation of the PDP in successive week-long visits to the WHAM device.  There was $\sim$1 visit each month for $\sim$6 months, starting with bare-floor and culminating in the successful TS and OES measurements reported here.  Between visits, time was utilized to order hardware and fabricate necessary parts, as well as to complete necessary administrative tasks, such as establishing approved laser-safety protocols.  ORNL personnel benefited from extensive support by Realta Fusion and UW-Madison personnel, that were present onsite at the WHAM facility.  Coordination with machine operations and vacuum breaks also influenced the work schedule.

\section{Experimental Results\label{sec:results}}

\begin{figure}[tbp]
    \centering
    \includegraphics[width=0.8\linewidth]{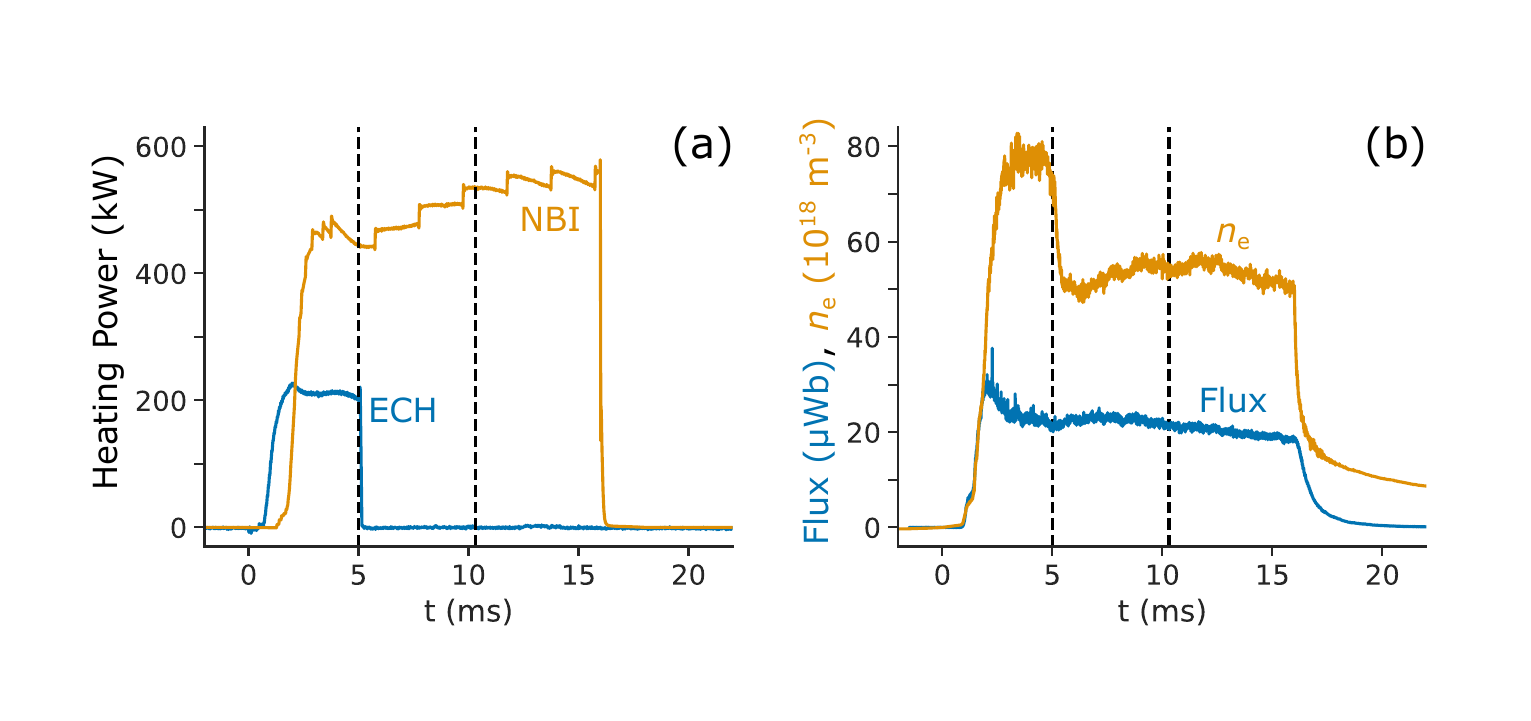}
    \caption{
    Overview of shot 250306075, a representative shot in the moderate density $\bar{n_e} = 5.0\times 10^{19}\ {\rm m}^{-3}$ ensemble. The OES system acquired at 5.0 ms, while the TS laser fired at 10.3 ms into the shot, as shown by the vertical dashed lines. (a) The ECH and NBI heating powers. (b) the averaged density from the line-integrated interferometer and the diamagnetic flux measurement from the fluxloop FL1 located at $z=7$ cm. 
    }
    \label{fig:shotsummary}
\end{figure}

The measurement capabilities of the PDP system are routinely used on WHAM plasmas.
A summary of the experiment for shot 250306075 is shown in \fref{fig:shotsummary}.
The plasma is generated by electron cyclotron heating (ECH) from $t$ = 0--5\,ms, while the neutral beam (NB) is injected from $t$ = 2--17\,ms.
The line-averaged electron density is measured by an interferometer, and adjusted by the calculated plasma diameter to infer an average electron density of $\sim$\,$5\times 10^{19}\mathrm{\;m^{-3}}$ (the blue curve in \fref{fig:shotsummary} right).
In \fref{fig:shotsummary}, the plasma flux measured by the flux loop is also shown by an orange curve, which with minimal assumptions can be converted to a total stored energy. 
From the line-averaged electron density and the stored energy, the average energy of a charged particle is estimated to be $\sim$\,50\,eV, by assuming that the same energy is distributed to the electrons and ions.

The laser for the Thomson scattering was injected at $t$ = 10.3\,ms, while the exposure time of the OES was made from $t$ = 5--5.1\,ms.
These timings are shown in the figure by vertical lines.

\subsection{OES Results}

\begin{figure}[tbp]
    \centering
    \includegraphics[width=0.4\linewidth]{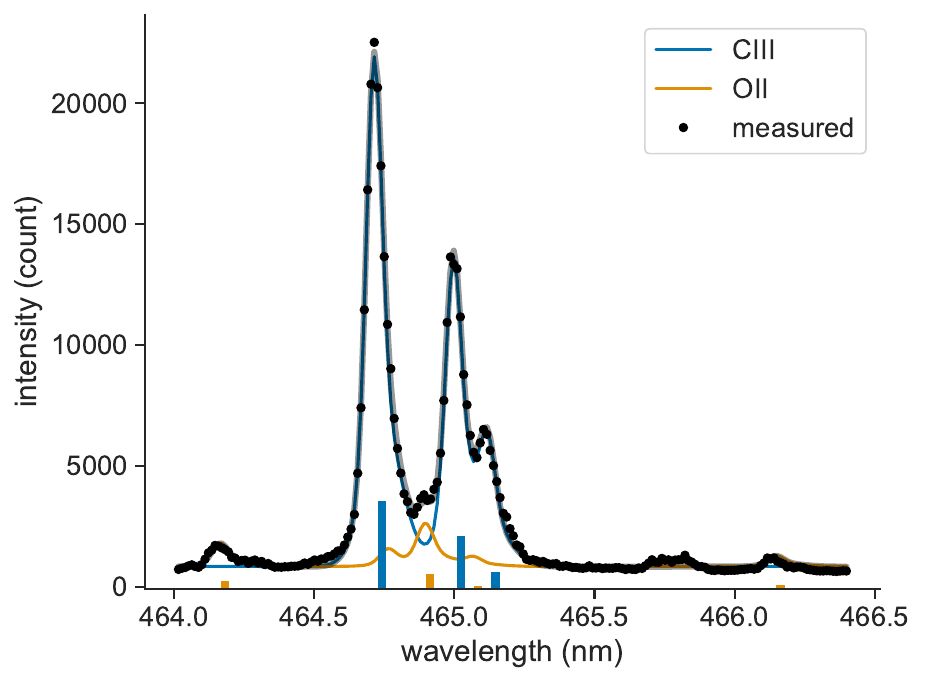}
    \caption{%
    Typical OES spectrum around 464--466 nm, where the black dots are the measured data and the gray solid line is the overall fit. The three prominent peaks arise from doubly charged carbon ions (blue vertical bars indicate their rest wavelengths and relative intensities), while the smaller features correspond to singly charged oxygen lines (orange). The spectral profile is obtained by convolving the Doppler broadening from a shifted Maxwellian ion velocity distribution with the asymmetric instrumental function (approximated by a sum of two Gaussians). Fitting this model to the observed spectrum yields the ion density, velocity, and temperature.
    }
    \label{fig:OES_spectrum}
\end{figure}

\Fref{fig:OES_spectrum} shows a typical spectrum obtained by the OES system.
The three big peaks are the emission lines from doubly charged carbon ions, i.e.,
464.742-nm line ($1s^22s3s\;^3\mathrm{S}_1 \leftarrow 1s^22s3p\;\mathrm{^3P^o}_2$),
465.025-nm line ($1s^22s3s\;^3\mathrm{S}_1 \leftarrow 1s^22s3p\;\mathrm{^3P^o}_1$),
and 
465.147-nm line ($1s^22s3s\;^3\mathrm{S}_1 \leftarrow 1s^22s3p\;\mathrm{^3P^o}_0$).
The rest wavelengths and their relative intensities (assuming a thermally distributed upper state population) are shown by blue vertical bars in the figure.
Also, singly charged oxygen lines are also visible at 
463.886 nm $2s^22p^2(\mathrm{^3P})3s\;\mathrm{4P}_{1/2} \leftarrow 2s^22p^2(\mathrm{^3P})3p\;\mathrm{^4D^o}_{3/2}$, 
464.181 nm $2s^22p^2(\mathrm{^3P})3s\;\mathrm{4P}_{3/2} \leftarrow 2s^22p^2(\mathrm{^3P})3p\;\mathrm{^4D^o}_{5/2}$, 
464.913 nm $2s^22p^2(\mathrm{^3P})3s\;\mathrm{4P}_{5/2} \leftarrow 2s^22p^2(\mathrm{^3P})3p\;\mathrm{^4D^o}_{7/2}$, 
and 
464.084 nm $2s^22p^2(\mathrm{^3P})3s\;\mathrm{4P}_{1/2} \leftarrow 2s^22p^2(\mathrm{^3P})3p\;\mathrm{^4D^o}_{7/2}$.

We assume a shifted Maxwellian for the ion velocity distribution, and fit the spectrum to obtain the density, velocity and temperature.
Since the wavelength resolution of our OES spectrometer ($\sim$\,0.04\,nm of the full-width-half-maximum, FWHM) is close to the ion Doppler width (the Doppler broadening of the 10\,eV carbon ions will be 0.03\,nm FWHM), an accurate evaluation of the instrumental profile is essential.
The instrumental profiles of the OES spectrometer has asymmetric shape, and thus we approximate it by a sum of two Gaussians (one with 0.04\,nm width and the other is 0.11\,nm width, 0.026\,nm apart, and with relative intensity difference of 0.5).
The spectral profile is obtained by convoluting the Doppler broadening to the instrumental function. 

These carbon and oxygen lines are multiplets. We assume a thermally distributed population in the upper state, and the relative intensity is obtained from the upper state population times the statistical weight and the Einstein A coefficient for each line.  The adjustable parameters for analysis of each spectrum are the upper state populations, shifts, and Doppler widths of the carbon and oxygen ions (totaling six) plus the continuum background emission level.  The fit results are shown by the gray solid curve in the figure. This reconstructs the observed spectrum, for the carbon (blue curve) and the oxygen (orange curve) contributions.

\begin{figure}[tbp]
    \centering
    \includegraphics[width=0.6\linewidth]{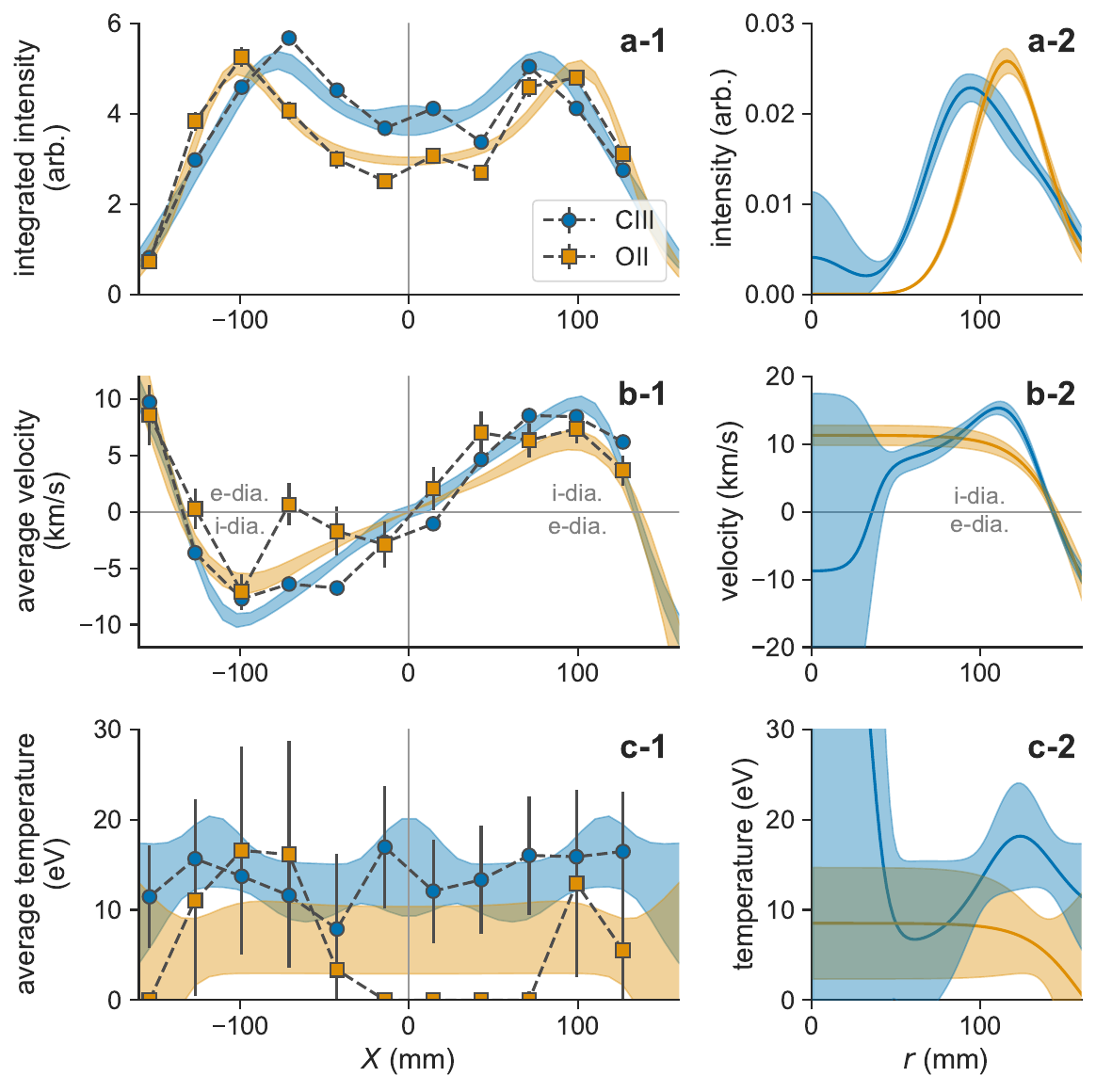}
    \caption{%
        Line-integrated (left column: a-1, b-1, c-1) and Abel-inverted local (right column: a-2, b-2, c-2) values of the upper-state population \(n_i\), ion velocity \(V_i\), and ion temperature \(T_i\) for doubly charged carbon (C\,III light) and singly charged oxygen (O\,II light). The local radial profiles obtained by Abel inversion (a-2, b-2, c-2) are shown with 1\(\sigma\) uncertainties (shaded bands), and reproduce the line-integrated data within their respective error bounds. The hollow radial profiles of \(n_i\) (a-2) show that the O\,II emission occurs at larger radii than the C\,III emission, consistent with the higher ionization energy needed for C\,III. The velocity profiles (b-2) have extrema around \(\lvert X\rvert \sim 100\) mm, with similar magnitudes for both ions, suggesting that the $E \times B$ drift is dominant over the diamagnetic drift. The ion temperatures (c-2) lie in the 10--20\,eV range, but its accuracy is limited by the spectrometer's instrumental width. 
    }
    \label{fig:OES_result}
\end{figure}

We repeat the same analysis for all eleven spectra from the available lines-of-sight. 
From the intensity, shift, and width, we derived the line-integrated values of the upper-state population $n_i$, ion velocity ($V_i$), and ion temperature $T_i$, which are shown in \fref{fig:OES_result} (a-1), (b-1), and (c-1), respectively.
Note that to cancel the mechanical drift of the spectrometer, we adopted the drift-correction described in the previous section.

The line-integrated values of $n_i$ are found to have a hollow shape for both of the ion species.
The emission location of the singly charged oxygen is more radially out-board than the doubly charged carbon ions. 
The line-integrated $V_i$ has an extrema around $|X|\sim80$\,mm, and becomes smaller in the outboard region of the plasma.
Furthermore, a similar velocity profile is found for the doubly charged carbon and singly charged oxygen ions.
The uncertainty in the $T_i$ result is $\sim$\,50\% and it is difficult to assess the spatial structure of $T_i$ at this stage of the diagnostic quality.

All the OES results are integrated along the sight lines. 
To assess the local value, we apply an Abel inversion for the observation results.
We assume that the zeroth, first, and second moments ($n_i$, $n_i \times V_i$, and $n_i \times T_i$) are axisymmetric, and can be described by a linear combination of Gaussian bases with a spatial extent of 50 mm.
The details of the inversion process are described in Appendix~\ref{app:abel}.

The local values of $n_i$, $V_i$, and $T_i$ assessed by the inversion process are shown in \fref{fig:OES_result}~(a-2), (b-2), and (c-2), respectively.
The shaded area indicates the 1-$\sigma$ uncertainty of the inversion.
The line-integrated values using these inversion results are shown in \fref{fig:OES_result}~(a-1), (b-1), and (c-1) with a $\pm \sigma$ uncertainty, which reasonably represents the measurement results.

The peak emission locations of C\,III and O\,II light are different: the O\,II is located more outboard.
As the ionization energy to generate $\mathrm{C}^{2+}$ ions (24.4 eV) is higher than that to generate $\mathrm{O}^{+}$ (13.6 eV), the emission location difference suggests the effect of the electron temperature gradient, i.e., the ion charge state increases towards the core.

The local distribution of $V_i$ for C\,III and O\,II are similar.
This velocity may include the diamagnetic drift due to the pressure gradient of these ions, and the $E \times B$ drift due to the potential gradient of the plasma.
Although the pressure gradient for these ions is not measurable, since $\mathrm{C^{2+}}$ and $\mathrm{O^+}$ are different in charge, the diamagnetic component of C\,III should be smaller than that of O\,II by a factor of 2.
On the other hand, the $E \times B$ drift component is the same for these ions.
The very similar velocity for these ions suggests that the $E \times B$ drift is dominant (over the diamagnetic drift) for the rotation.

The plasma potential has been measured at the end cell for a similar experiment. 
By assuming that the plasma potential is a function of the magnetic flux and constant along the magnetic field line (across the central chamber and the end cells), the electric field in the plasma has been evaluated.
The $E \times B$ evaluated from the end-cell potential indicates the $E \times B$ drift toward the ion-diamagnetic direction, the linear profile across the plasma radius, and its velocity with $10^4$\,m/s at $r=100$\,mm, all of which is consistent with our OES measurement.
This agreement also suggests that the ion velocity mainly comes from the $E \times B$ drift.

The ion temperature is in the range of 10--20\,eV. However, as the instrumental width is larger than the Doppler broadening, its accuracy is limited.
This makes the inversion of ion temperature more challenging than the other parameters. 

\subsection{TS results}

The TS system was used to measure electron parameters in the WHAM experiment. 
The data presented here are from two sets of commissioning shots on WHAM chosen to maximize the likelihood of making a successful TS measurement. This mode of operation, though useful for this testing of the TS diagnostic system, is not representative of high-performance plasma operations on WHAM. 
Two WHAM plasma settings were chosen for their repeatable density time-histories: a moderate density case of peak density near $5\times 10^{19}{\ \rm m}^{-3}$ (37 shots in 250306045-250306081, the shot presented in the previous subsection, \fref{fig:shotsummary}, is also included) and a high density case of above $1\times 10^{20}{\ \rm m}^{-3}$ (20 shots in 250305121-250305143).

\begin{figure}[tbp]
    \centering
    \includegraphics[width=0.6\linewidth]{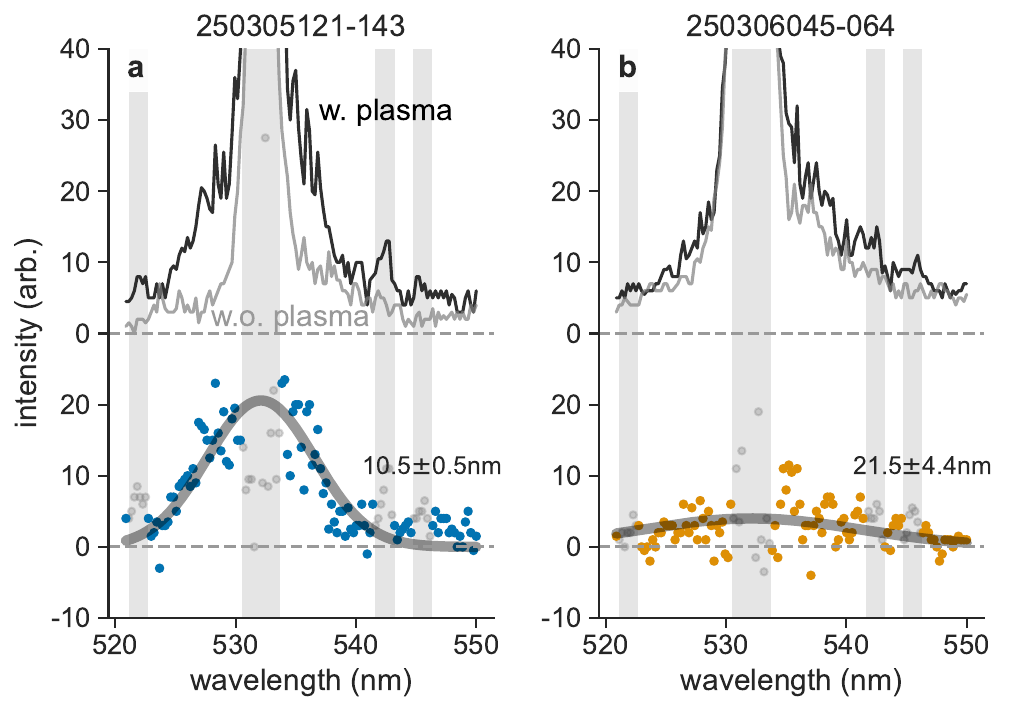}
    \caption{%
    Typical Thomson scattering (TS) spectra from two WHAM plasma experiments: (a) the high-density and (b) the moderate-density conditions. In each upper panel, the black trace shows the spectrum with the laser injection while the plasma is on, and the gray trace shows the spectrum after the plasma has disappeared. Multiple shots (20 for the high-density and 37 for the moderate-density conditions) are averaged to improve the signal-to-noise ratio. The lower panels display the differential spectra, revealing a Gaussian-like TS feature centered on the laser wavelength (with gray-shaded regions masked to remove impurity emission). The overlaid thick curves are best-fit Gaussians (centroid fixed at the laser wavelength, zero background) with the FWHM values indicated. As expected, the higher-density plasma exhibits a narrower, more intense TS signal, whereas the lower-density plasma shows a broader scattered profile.
    }
    \label{fig:TS_spectrum}
\end{figure}

\Fref{fig:TS_spectrum} shows typical TS spectrum obtained for these two cases. 
The ensembled spectra are shown in the upper panels, with the black lines collected during the plasma (t=10.3\,ms) and the gray lines obtained after the plasma (t=110.3\,ms).
In order to increase the signal-to-noise ratio (SNR), we repeated each shot 20 and 23 times and averaged the obtained spectra, for the two conditions, respectively. 
The differential spectra are shown in the lower panel, and for the higher-density experiment a clear difference can be seen. 
This differential spectra shows a Gaussian-like bell shape centered around the laser wavelength.

The differential spectra contain both the Thomson scattering light, as well as the plasma emission mainly due to impurity emissions.
We mask these regions, as indicated by the gray hatches in the figure.
The scattered signal from the higher density plasma is narrower and more intense, as expected.

We fit the unmasked spectrum with a Gaussian distribution, which has its centroid fixed to the laser wavelength and zero background, as shown by thick curves.
The optimal Full Width Half Max (FWHM) widths are also shown in the figure.
A broader width is found for the lower-density experimental conditions.

Since the laser wavelength is 532\,nm and we observe Thomson scattering light from the $90^\circ$ orientation, the value of FWHM and the electron temperature are related by:
\begin{align}
    \frac{T_\mathrm{e}}{\mathrm{eV}} \approx 0.163 \times \left(\frac{\mathrm{FWHM}}{\mathrm{nm}}\right)^2,
\end{align}
which assumes non-relativistic electrons.
From this relation, electron temperatures $T_e$ of $18\pm2$\,eV and $75\pm31$\,eV are measured for these two experiments, respectively.
From the intensity (area) of the TS spectrum, the electron density can be estimated.
For the two ensembles, electron densities $n_e$ of $(1.0\pm 0.1)\times 10^{20} \mathrm{\;m^{-3}}$ and $(5.1\pm 0.8)\times 10^{19} \mathrm{\;m^{-3}}$ are measured, respectively.

\begin{figure}[tbp]
    \centering
    \includegraphics[width=0.4\linewidth]{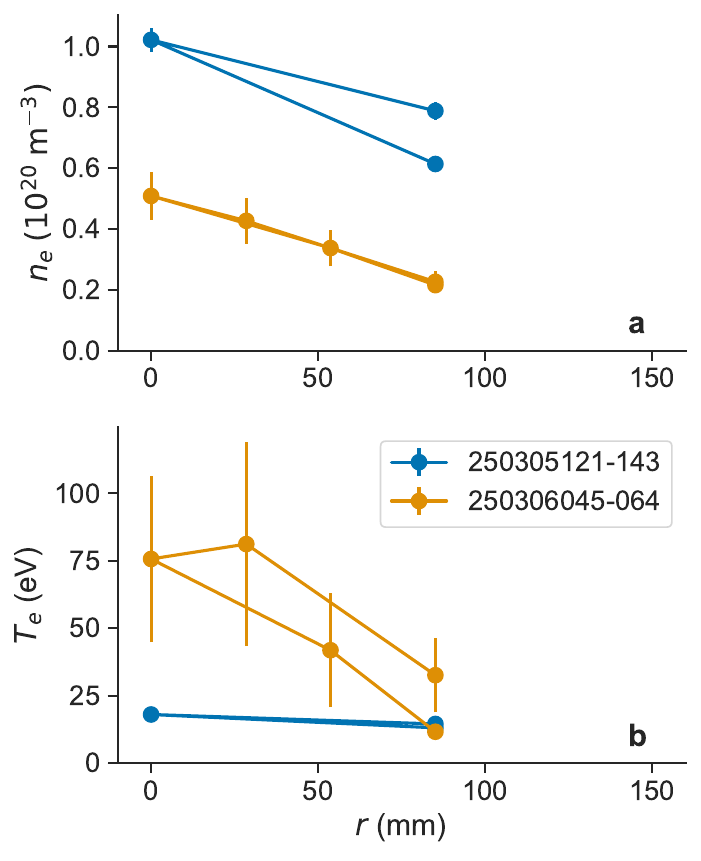}
    \caption{%
        Radial profiles of (a) \(n_e\) and (b) \(T_e\)  from Thomson scattering measurements for the high and low density WHAM ensembles. The densities agree well with the line-integrated interferometric measurements, and the centrally peaked \(T_e\) reaches $\sim$\,70\,eV for the lower-density case. Around \(r \sim 70\)\,mm, where \(\mathrm{C}^{2+}\) is depleted, \(T_e \sim 50\)\,eV. Error bars represent the statistical uncertainties from repeated shots and the fit procedure.
        The lines connect the measurement points for clarity.
        Note that while five measurement points are available for 250306045-064, due to one broken fiber and too-strong stray-light for the two sightlines at both the edges only three sightlines were utilized for 250305121-143.
    }
    \label{fig:TS_result}
\end{figure}

We repeated the same analysis for other sight lines. As shown in \fref{fig:TS_result}, both the profiles of $n_e$ and $T_e$ are centrally peaked.

The values of $n_e$ are consistent with the sight-line averaged density measured by the interferometer, for both cases. 
The central $T_e$ is as high as $\sim$\,70\,eV, although the SNR of the $T_e$ profile is still low. 
It is also suggested that the place where $\mathrm{C}^{2+}$ is depleted $r\sim$ 70\,mm, which can be seen in \fref{fig:OES_result}, has $T_e \sim 50$\,eV.
This is consistent with the $T_e$ dependence of the ionization rate of $\mathrm{C}^{2+}$, which has 47.9\,eV ionization potential.

\section{Summary and future work\label{sec:summary}}

In this work, we have demonstrated the application of the PDP to WHAM.
The PDP integrates an active TS system and an OES system into a single, compact unit that can be rapidly deployed and integrated into existing plasma experiments.
The OES system facilitates a comprehensive impurity line survey and enables flow measurements through the Doppler effect observed in impurity lines.
Notably, plasma rotation profiles were successfully derived from doubly charged carbon lines, underscoring the diagnostic package's potential for advancing experimental plasma studies.
The TS system enables the first-ever measurement of the electron temperature distribution in WHAM, offering critical insights into plasma behavior.
We reiterate that these shots on WHAM are commissioning shots chosen for their high density and repeatable behavior, and not for high performance. 

There are several areas where the system can be improved.
Firstly, the inversion accuracy of the OES system is limited by the number of sight lines. 
With more sight lines, particularly in the edge region, the velocity shear may be more accurately inferred.
Secondly, the SNR of the TS system is limited by the contrast of the TS signal against the stray light and plasma background signal.
Using bigger collection optics, higher-NA optical fibers, and higher F/\# spectrometer will improve the SNR.
The SNR is currently limited by the photon noise of the stray light, not the photon noise of the scattering light. 
By using 0.39 NA fiber (and with matched fiber connections and spectrometers) we may be able to increase the throughput by a factor of 20x (2x by properly matching the fiber connection and 10x by increasing the NA).
An increase in throughput by a factor of 20 increases the SNR to a level equivalent to performing a 400 experiment ensemble. 
The experiment where the TS measurement is demonstrated here has much higher electron-density and lower electron temperature than a typical high-performance WHAM plasma.
The improvement may enable a single-shot TS measurement by the PDP for such high-performance WHAM plasmas.

\appendix

\section{Time sequence to control TS timing}\label{app:timing}

\begin{figure}[tbp]
    \centering
    \includegraphics[width=0.9\linewidth]{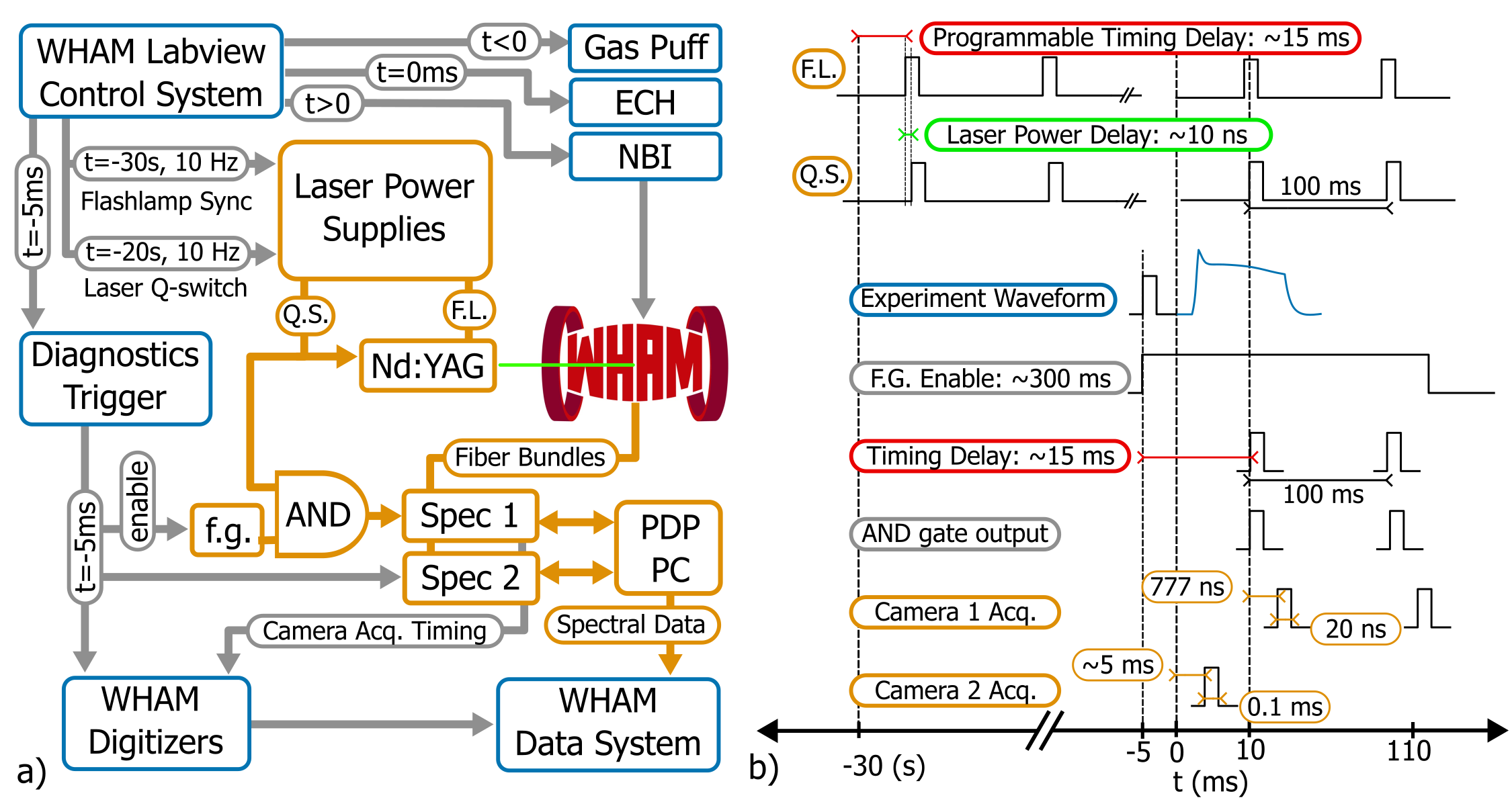}
    \caption{%
        Timing diagram for implementation of the PDP on WHAM. (a)  Process flow diagram of interfaces between WHAM systems (blue) and ORNL systems (orange), from the WHAM Control System to the WHAM Data Server.  (b) Representative timing diagram (not to scale), beginning at $t=-30$\,s to synchronize the PDP with the plasma pulse ($t\sim 0$), including 1) TS camera system and laser pulses on $\sim$\,ns time scales, and 2) OES camera system on $\sim$\,ms time scales.
    }
    \label{fig:WHAM_PDP_timing}
\end{figure}

The timing sequence for implementing the PDP on the WHAM system is crucial for ensuring the proper synchronization of diagnostic components with plasma production. A particular challenge is the synchronization of the laser pulse, camera exposure, and plasma pulse, as the laser pulse duration is much shorter ($\sim$\,10 ns) compared to the plasma pulse ($\sim$\,10 ms), compared to machine system process logic controls ($\sim$\,10 s). Minimizing the camera exposure time is also essential to reduce the background emission from the plasma, improving the SNR. Furthermore, the laser must be operated a few seconds before the plasma pulse to allow for laser stabilization.

The process flow diagram on the left side of Figure~\ref{fig:WHAM_PDP_timing} illustrates the interfaces between the WHAM systems (depicted in blue) and the ORNL systems (shown in orange), outlining the communication and control pathways from the WHAM Control System to the WHAM Data Server. The WHAM Control System sends a trigger signal to the laser 30 seconds before the plasma pulse. This timing is adjusted up to 10 ms to control the measurement timing during the plasma pulse. 

After receiving the trigger signal, the flashlamps in the laser start flashing at a 10 Hz repetition rate. Due to limitations in the laser control software and safety requirements, a human operator must initiate the laser Q-switching, which is done approximately 10 seconds before the plasma pulse.

The TS spectrometer camera is synchronized with the laser Q-switch. To identify the single laser pulse during the plasma experiment (as well as the subsequent pulse for stray light measurement), we use an AND logic gate, which combines the Q-switch synchronization signal generated by the laser and a signal indicating plasma operation (generated by a function generator in response to the WHAM Diagnostic trigger system at $t=-5$\,ms, and extended to TTL high for 250 ms).

The OES camera is triggered by the WHAM diagnostic trigger, which is sent to the camera 5 ms before the plasma pulse. The trigger delay is fine-tuned using the camera control software. After acquiring the data, the camera signals are transmitted to the WHAM Data Server.

\section{Rayleigh scattering calibration of the TS system}\label{app:RayleighCalibration}

\begin{figure}[tbp]
    \centering
    \includegraphics[width=1.0\linewidth]{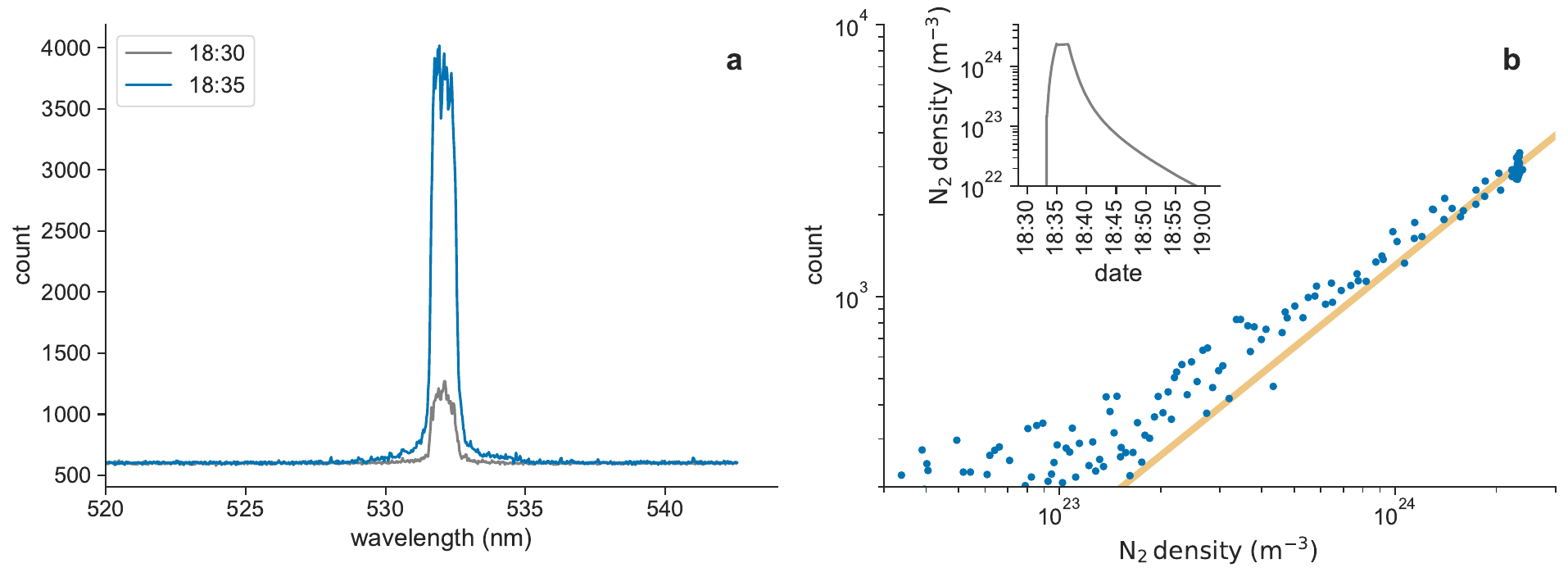}
    \caption{%
    Calibration of the Thomson scattering (TS) system using Rayleigh scattering from nitrogen gas. (a) Spectral measurements before and after nitrogen gas injection. The grey spectrum (18:30) shows stray light from the laser, while the blue spectrum (18:35) shows a significant increase in intensity due to Rayleigh scattering from the nitrogen gas. (b) 
    A correlation between the nitrogen gas density measured measured by the pressure gauge and the Rayleigh scattering intensity.
    The solid line shows its linear fit.
    Temporal evolution of the nitrogen gas density is shown in the inset.
    }
    \label{fig:rayleigh}
\end{figure}

The sensitivity of the Thomson scattering (TS) system is calibrated using Rayleigh scattering from a nitrogen gas backfill, consistent with techniques reported in Ref.~\cite{LeBlanc2008-yy}. 
The inset of \fref{fig:rayleigh}~(b) shows the nitrogen gas density, as calculated using the ideal gas law from the measured total pressure gauge readings. The gas is injected into the vacuum chamber at 18:33, and the pressure is maintained for several minutes before being pumped out.

Rayleigh scattering light is detected using the TS system hardware. Since the Rayleigh light shares the same wavelength as the laser light, we adjusted the spectrometer's grating angle to focus the laser spectrum onto a portion of the chip positioned away from the optical block. The observed spectrum before and after the gas injection is shown in \fref{fig:rayleigh}~(a). Prior to gas injection, stray light from the laser is evident. After the injection, the spectral intensity increases significantly due to the Rayleigh scattering from the nitrogen gas.

We integrate the spectrum to obtain the total intensity of the Rayleigh scattering light and stray light background. 
\Fref{fig:rayleigh}~(b) shows the relation of Rayleigh scattering intensity and the nitrogen gas density, $n_g$.
These two signals show a good proportionality. This proportional coefficient, \(C_R\), is used to determine the sensitivity of the TS system.  The average of the light intensity data below $n_g\sim10^{23}$ in this plot can be used to estimate the stray light background intensity, \(I_{SLB}\).  The measured signal during Rayleigh scattering backfill of nitrogen is the sum of the Rayleigh scattered photons and the stray light background photons, \(I_{m,RS}=I_R+I_{SLB}\).  In general, the Rayleigh scattered photon intensity is proportional to the cross-section for Rayleigh scattering \(\sigma_R\), the nitrogen gas density \(n_g\), the laser energy \(E_l\), the detector gain \(G_R\), and a coefficient representing optical collection geometry \(C_{opt}\):  \(I_R=\sigma_R n_g E_l G_R C_{opt}\).  Similarly, the measured signal during Thomson scattering is the sum of the Thomson scattered photons and the stray light background photons, \(I_{m,TS}=I_T+I_{SLB}\).  And, the Thomson scattered photon intensity is proportional to the cross-section for Thomson scattering \(\sigma_T\), the plasma electron density \(n_e\), the laser energy \(E_l\), the detector gain \(G_T\), and a coefficient representing optical collection geometry \(C_{opt}\):  \(I_T=\sigma_T n_e E_l G_T C_{opt}\).  Taking the ratio of \(I_T/I_R\) and rearranging eliminates the unmeasured optical constants, the detector gains (if set equal during Rayleigh and Thomson measurements) and the laser energy (if roughly constant).

Thus, using the Rayleigh scattering calibration light, we can estimate the electron density from the TS spectral intensity. The electron density, \( n_e \), can be derived from the TS light intensity, \( I_T \), and the Rayleigh scattering light intensity, \( I_R \), using the following equation:
\begin{align}
    n_e \approx 0.029\,n_g\,\frac{I_T}{I_R} \approx 0.029\,\frac{I_T}{C_R},
\end{align}
where $n_g$ is the nitrogen gas density, and $C_R$ is the proportional coefficient (the slope of the line in \Fref{fig:rayleigh}~(b)).~\cite{Barth2001-kt} In this calibration, we assume the use of a 532\,nm laser and a 90-degree scattering angle.

\section{Abel inversion technique for the OES measurement}\label{app:abel}

The Abel inversion technique is used to derive local values from the line-integrated values obtained by the OES system. The method employed in this study combines a Gaussian basis expansion with basis selection using the nonnegative least squares method.

Initially, we assume that the local value can be expressed as a linear combination of Gaussian basis functions:
\begin{align}
    f(r) = \sum_{i=1}^{N} a_i \phi_i(r),
    \label{eq:gaussian_basis}
\end{align}
where
\begin{align}
    \phi_i(r) = \exp\left(-\frac{(r - r_i)^2}{2\sigma^2}\right),
\end{align}
with \(a_i\) representing the linear coefficients, \(r_i\) the center of the \(i\)-th Gaussian basis, and \(\sigma\) the width of the Gaussian function. In this study, \(\sigma = 20\) mm, and the interval for \(r_i\) is 20 mm.

Next, we calculate the path length for each basis along each sight line. Since the light intensity and other moments are integrated along the sight line, the line-integrated value can be expressed as:
\begin{align}
    I_j = \sum_{i=1}^{N} a_i A_{ij},
\end{align}
where
\begin{align}
    A_{ij} = \int dL_j \, \phi_i(r).
\end{align}
Here, \(\int dL_j\) denotes the integration along the \(j\)-th sight line.

To obtain the local values from the line-integrated intensities, we solve for \(a_i\) that minimizes the difference between the measured line-integrated intensity and the calculated value, based on equation~\ref{eq:gaussian_basis}. A key challenge of this inversion method lies in the non-uniqueness of the solution. In particular, since emission predominantly originates from the plasma edge region, the local value in the core region may not be well constrained.

To address this issue, we select the most relevant basis functions from the intensity data (zeroth moment), assuming nonnegativity for the coefficients \(a_i\). To find the optimal values of \(a_i \geq 0\), we employ the nonnegative least squares method. Due to the nonnegative constraint, some of the \(a_i\) values may be zero, implying that the corresponding Gaussian basis functions are not included in the expansion.

The same set of basis functions is applied for calculating the first and second moments. For the first moments, the nonnegativity constraint is relaxed, and the least squares method is used to find the optimal values of \(a_i\).

\bibliography{refs}

\begin{acknowledgments}
    This work was supported by the U.S. D.O.E contract DE-AC05-00OR22725, and funded under the INFUSE program -- a DOE SC FES public-private partnership --  under CRADA No. NFE-24-10027 between Oak Ridge National Laboratory and Realta Fusion Inc.
\end{acknowledgments}


\begin{widetext}
Notice:  This manuscript has been authored by UT-Battelle, LLC, under contract DE-AC05-00OR22725 with the US Department of Energy (DOE). The US government retains and the publisher, by accepting the article for publication, acknowledges that the US government retains a nonexclusive, paid-up, irrevocable, worldwide license to publish or reproduce the published form of this manuscript, or allow others to do so, for US government purposes. DOE will provide public access to these results of federally sponsored research in accordance with the DOE Public Access Plan (\url{http://energy.gov/downloads/doe-public-access-plan}).
\end{widetext}

\end{document}